\documentclass[12pt]{article}

\global\arraycolsep=1pt

\usepackage{amsbsy,amssymb,latexsym,amsfonts, amsmath}
\usepackage{mathrsfs}
\usepackage{graphicx}
\usepackage{cite}
\usepackage{bm}
\usepackage{here}

\numberwithin{equation}{section}
\newcommand{\bel}[1]{\begin{equation}\label{#1}}                     
\newcommand{\bal}[1]{\begin{eqnarray}\label{#1}}                     
\newcommand{\be}{\begin{equation}}
\newcommand{\ee}{\end{equation}}

\newcommand{\im}{\mathrm{i}}
\newcommand{\ex}{\mathrm{e}}
\newcommand{\de}{\mathrm{d}}

\renewcommand{\thefootnote}{\fnsymbol{footnote}}
\newcommand{\bea}{\begin{equation}}
\newcommand{\eea}{\end{equation}}
\newcommand{\bra}[1]{\langle{#1}|}
\newcommand{\ket}[1]{|{#1}\rangle}

\def\CH{{\mathcal H}}

\def\CN{\mathcal N}

\def\CK{\mathcal{K}}

\begin{document}
%%%%%%%%%%%%%%%%%%%%%%%%%%%%%%%%%%%%%%%%%%%%%%%%%%%%%%%%%%%%%%%%%%%%%%%%%%%%%%%%%%%%%%%%%%
%
% title page
%
%%%%%%%%%%%%%%%%%%%%%%%%%%%%%%%%%%%%%%%%%%%%%%%%%%%%%%%%%%%%%%%%%%%%%%%%%%%%%%%%%%%%%%%%%
\begin{titlepage}
%%%%%%%%%%%%%%%%%%%% preprint # %%%%%%%%%%%%%%%%%%
\begin{flushright}
\normalsize
%\filename
~~~~
OCU-PHYS 436\\
December, 2015\\
\end{flushright}
%%%%%%%%%%%%%%%%%%%%%%%%%%%%%%%%%%%%%%%%%%%%%%%%%%

\vspace{15pt}

%%%%%%%%%%%%%%%%%%%% title %%%%%%%%%%%%%%%%%%%%%%%
\begin{center}
{\LARGE The integral representation of solutions of KZ equation } \\
\vspace{5pt}
{\LARGE and a modification by  $\CK$ operator insertion}\\
\end{center}
%%%%%%%%%%%%%%%%%%%%%%%%%%%%%%%%%%%%%%%%%%%%%%%%%%

\vspace{23pt}

%%%%%%%%%%%%%%%%%%% authors %%%%%%%%%%%%%%%%%%%%%%
\begin{center}
{Reiji Yoshioka\footnote{e-mail yoshioka@sci.osaka-cu.ac.jp}  }\\
%%%%%%%%%%%%%%%%%%%%%%%%%%%%%%%%%%%%%%%%%%%%%%%%%%
%
\vspace{18pt}
%
%%%%%%%%%%%%%%%%%%% affiliation %%%%%%%%%%%%%%%%%%%

\it Osaka City University Advanced Mathematical Institute (OCAMI)

\vspace{5pt}

3-3-138, Sugimoto, Sumiyoshi-ku, Osaka, 558-8585, Japan \\

\end{center}
%%%%%%%%%%%%%%%%%%%%%%%%%%%%%%%%%%%%%%%%%%%%%%%%%%%
%
\vspace{20pt}
\begin{center}
Abstract\\
\end{center}
%%%%%%%%%%%%%%%%%%%% abstract %%%%%%%%%%%%%%%%%%%%%
A root of unity limit of the $q$-deformed Virasoro algebra is considered. 
The $\widehat{sl}(2)_k$ current algebra and the integral formulas of the solutions of the KZ equations 
 can be realized by the $q$-deformed boson at the limit and an additional boson. 
We explicitly construct the integral representation of the four-point blocks 
 with a $\CK$-operator insertion. 
%%%%%%%%%%%%%%%%%%%%%%%%%%%%%%%%%%%%%%%%%%%%%%%%%%%

\vfill

\setcounter{footnote}{0}
\renewcommand{\thefootnote}{\arabic{footnote}}

\end{titlepage}

%%%%%%%%%%%%%%%%%%%%
\renewcommand{\thefootnote}{\arabic{footnote}}
\setcounter{footnote}{0}
%%%%%%%%%%%%%%%%%%%%

%%%%%%%%%%%%%%%%%%%%%%%%%%%%%%%%%%%%%%%%%%%%%%%%%%%%%%%%%%%%%%%%%%%%%%%%%%%%%%%%%%%%%%%%%%%%%%%%%%%% section 1
\section{Introduction}
\label{sec:intro}
The AGT relation \cite{AGT} states that the instanton partition functions \cite{Nekrasov}
  of the four-dimensional $\mathcal{N}=2$ $SU(2)$ gauge theory 
  are related to the correlation functions \cite{BPZ} of the 
 two-dimensional conformal field theory with Virasoro symmetry.  
The extension to the similar correspondence between $SU(n)$ gauge theory and 
 conformal field theory with $W_n$ symmetry has been constructed in \cite{Wyl0907, MM0908}. 
Since then, the both sides of the correspondence have been intensively studied 
 by a number of people. 
For example, see \cite{DV,IMO,MMS0911,FL0912,MMS1001,IO5,MMM1003,IOYone,MMS1012,BMT1104,KMZ1306,
MS1307,MMZ1312,MRZ1405,BF1105,NT1106,BMT1106,BMT1107,BBB1106,Wyll1109,Ito1110,AT1110,BW1205,BM1210}. 

There exists a natural generalization to 
 the connection between the 2d theory with the $q$-deformed Virasoro/W symmetry 
 and the five-dimensional gauge theory \cite{AY0910,AY1004,MMSS1105,AFHKSY1106,Ohku1404,Zenk1412,MZ1510}.  
Recently, the elliptic Virasoro/6d correspondence is also proposed in \cite{IKY1511,Nier1511}.

In our previous papers \cite{IOY2,IOY3,IOYconf}, 
 we considered a $r$-th root of unity limit in $q$ and $t$ 
 ($q \to \omega$, $t \to \omega$ with $\omega= \ex^{\frac{2 \pi \im}{r}}$)
 of the $q$-W/5d correspondence.
In \cite{IOY2}, we proposed a limiting procedure to 
 get the Virasoro/W block in the 2d side from that in the $q$-deformed version. 
In \cite{IOY3}, we have elaborated the limiting procedure and 
 showed that the ${\bf Z}_r$-parafermionic CFT appears in the 2d side.
\footnote{The root of unity limit of the troidal algebra has 
 also been considered in \cite{BBT1211B}. }

The another extension of the AGT relation by including various defects 
 in the gauge theory have also been considered \cite{AGGTV}.   
We are interested in the defects, the so-called surface operator, 
 supported on two-dimensional submanifolds 
 (for review for surface operator, see \cite{Guko1412}). 
There are two kinds of the surface operators in a sense, i.e.  
 two-dimensional defects brought either from 2d-defect of 4d-defects
 in the M5-brane construction of $\CN = 2$ gauge theory. 
The instanton partition functions in the presence of 
 a kind of surface operator are related with 
 the conformal block with a degenerate field insertion 
 \cite{KPW1004,DGH1006,MT1006,Taki1007,Gaio0908A,MMM0909O,AFKMY1008}.  
In the present paper, we focus on the 4d-defects. 
It was conjectured that the instanton partition functions in this case
 are related to the affine $\widehat{sl}_k(n)$ current block with a mysterious operator, 
 the so-called $\CK$ operator, insertion \cite{AT1005,KPPW1008}. 
More general cases of the surface operator insertion are discussed in \cite{Wyll1011,Wyll1012}. 
 
On the other hand, 
 the defining relation of the $q$-Virasoro algebra is  \cite{SKAO,FF95,AKOS95} 
\be \label{q-virasoro}
f(z'/z) \mathcal{T}(z) \mathcal{T}(z') 
- f(z/z') \mathcal{T}(z') \mathcal{T}(z)
= \frac{(1-q)(1-t^{-1})}{(1-p)}
\Bigl[ \delta(pz/z') - \delta(p^{-1}z/z') \Bigr],
\ee
where $p = q/t$ and 
\be
f(z) = \exp\left( \sum_{n=1}^{\infty} \frac{1}{n}
\frac{(1-q^n)(1-t^{-n})}{(1+p^n)} z^n \right), ~~~~~
\delta(z) = \sum_{n \in \mathbb{Z}} z^n.  
\ee
It is known that the $q$-deformed current $\mathcal{T}(z)$ 
 can be realized by the $q$-deformed Heisenberg algebra. 
In this paper, we consider the following root of unity limit in $q$ and $t$:  
\be
 q \to 1, ~~~ t \to -1.  
\ee
This limit has been considered in \cite{HJKO0005O} and 
 the $q$-Virasoro algebra \eqref{q-virasoro} is reduced to 
  the Lepowsky-Wilson's $Z$-algebra \cite{LW1981C,LW1981A}. 
The $\widehat{sl}(2)_k$ current algebra can be realized 
 by using the two types of bosons obtained from a $q$-deformed boson in this limit 
 and an additional boson.
In this formalism, we will first reconstruct the integral representation of
 the solution to the KZ equation \cite{KZ} that the current blocks should satisfy. 
Then we consider the current block with the $\CK$ operator insertion 
 and illustrate how to derive those as integral representation.

This paper is organized as follows: 
 In section two, the root of unity limit of the $q$-deformed boson is considered. 
 In section three,  we see that the $\widehat{sl}(2)_k$ current algebra can be realized. 
 In section four, integral formulas of the solutions to  
  the KZ equation are constructed.
 In section five, we consider the four-point current block
 with the $\CK$ operator insertion and present its integral representation.   
%%%%%%%%%%%%%%%%%%%%%%%%%%%%%%%%%%%%%%%%%%%%%%%%%%%%%%%%%%%%%%%%%%%%%%%%%%%%%%%%%
%%%%%%%%%%%%%%%%%%%%%%%%%%%%%%%%%%%%%%%%%%%%%%%%%%%%%%%%%%%%%%%%%%%%%%%%%%%%%%%%%
\section{$\bm{q}$-deformed boson at root of unity}
%%%%%%%%%%%%%%%%%%%%%%%%%%%%%%%%%%%%%%%%%%%%%%%%%%%%%%%%%%%%%%%%%%%%%%%%%%%%%%%%%
As mentioned in Introduction, the $q$-deformed Virasoro current $\mathcal{T}(z)$
 can be realized  by the $q$-deformed Heisenberg algebra \cite{SKAO}, 
\be
\begin{split}
 &[\alpha_0,\widetilde{Q}]=2, \\
 &[\alpha_n,\alpha_m] =  -\frac{1}{n}(1-q^{-n})(1-t^{n})
 (1+p^{n}) \delta_{n+m,0}. 
\end{split} 
\ee
The $q$-deformed boson is defined by 
\begin{align}
 \widetilde{\varphi}(z)
 = \widetilde{\varphi}_{\rm even}(z) + \widetilde{\varphi}_{\rm odd}(z), ~~~~
\end{align}
where 
\begin{align}
 \widetilde{\varphi}_{\rm even}(z) &= 
 \beta^{\frac{1}{2}} \widetilde{Q} 
 + \beta^{\frac{1}{2}} \alpha_0 \log z 
 - \sum_{n \neq 0} \frac{1}{q^{n} - q^{-n}} 
 \alpha_{2n} z^{-2n}, \\
 \widetilde{\varphi}_{\rm odd} (z) &= 
 -\sum_{n \in {\bf Z}} 
 \frac{1}{q^{(2n+1)/2} - q^{-(2n+1)/2}} 
 \alpha_{2n+1} z^{-(2n+1)}.  
\end{align}
Let us consider the following limit of the $q$-deformed boson, 
\be
 q = t^{\alpha}= \ex^{-\frac{h}{\sqrt{\beta}}}, ~~~~ 
 t  = (-1)^{\tilde{k}}  \ex^{-\sqrt{\beta} h}, ~~~~ 
 p = q/t = (-1)^{-\tilde{k}} \ex^{Q_E h}, ~~~~
 h \to 0. 
 \label{limit:q=1,t=-1}
\ee
where $\tilde{k}$ is an odd number and $Q_{E} = \sqrt{\beta} - \frac{1}{\sqrt{\beta}}$ and 
$\alpha = \frac{1}{\beta}$. 
In order to take this limit, the following condition is demanded:   
\be
 \alpha \left(\im \pi k - \frac{h}{\sqrt{\alpha}} +  2 \pi \im m_+ \right)
 = - \sqrt{\alpha} h + 2 \pi \im m_-. 
\ee
where $m_{\pm}$ are positive integers. 
Thus the parameter $\alpha$ has to satisfy 
\be
 \alpha = \frac{1}{\beta} = \frac{2m_-}{2m_+ + \tilde{k}} \equiv \frac{2}{2 + k}, 
\ee
where
\be
 k = \frac{2(m_+ - m_-) + \tilde{k}}{m_-}. 
\ee
In the section 3, 
 we will see that the parameter $k$ serves as the level of 
 the $\widehat{sl}(2)_k$ algebra.
%From now on we set $k>0$.  
In the limit ({\ref{limit:q=1,t=-1}), 
 the even and odd part of the $q$-boson are expanded in powers of $h$ as  
\begin{align} 
  \widetilde{\varphi}_{\rm even}(z) 
 %= \sqrt{\beta} \left\{ Q + \alpha_0 \log z 
 %- \sum_{n \neq 0}  \frac{a_{2n}}{2n} z^{-2n} \right\} 
 %+ \mathcal{O}(h) 
 &\equiv \phi_1(z) + \mathcal{O}(h), \\
  \widetilde{\varphi}_{\rm odd} (z) 
 %= \sqrt{\beta-1}  
 %\left[- \sum_{n \in {\bf Z}} \frac{a_{2n+1}}{2n+1} z^{-(2n+1)} \right]
 %+ \mathcal{O}(h) 
 &\equiv \phi_2(z) + \mathcal{O}(h), 
\end{align} 
where 
\begin{align}
\phi_1(z) & %= \sqrt{\beta}\phi(z) 
 = Q + \phi^{(1)}_{0} \log z 
 - \sum_{n \neq 0} \frac{\phi^{(1)}_{2n}}{2n} z^{-2n}, \\
 \phi_2(z) & %= \sqrt{\beta-1} \phi(z) 
 = - \sum_{n \in \textbf{Z}} \frac{\phi^{(2)}_{2n+1}}{2n+1} z^{-(2n+1)}. 
\end{align}
with the commutation relations, 
\begin{align}
 &[\phi^{(1)}_{2n} , \phi^{(1)}_{2m}] = (k+2) (2n) \delta_{n+m,0}, ~~~~~
 [\phi^{(1)}_{0} , Q ]=  k+2, \\
 &[\phi^{(2)}_{2n+1} , \phi^{(2)}_{-2m-1}] = -k(2n+1) \delta_{n,m}. 
\end{align}

%%%%%%%%%%%%%%%%%%%%%%%%%%%%%%%%%%%%%%%%%%%%%%%%%%%%%%%%%%%%%%%%%%%%%%%%%%%%%%%%%
%%%%%%%%%%%%%%%%%%%%%%%%%%%%%%%%%%%%%%%%%%%%%%%%%%%%%%%%%%%%%%%%%%%%%%%%%%%%%%%%%
\section{Affine $\bm{\widehat{sl}_k(2)}$ algebra}
%%%%%%%%%%%%%%%%%%%%%%%%%%%%%%%%%%%%%%%%%%%%%%%%%%%%%%%%%%%%%%%%%%%%%%%%%%%%%%%%%
%%%%%%%%%%%%%%%%%%%%%%%%%%%%%%%%%%%%%%%%%%%%%%%%%%%%%%%%%%%%%%%%%%%%%%%%%%%%%%%%%
\subsection{Free field realization}
%%%%%%%%%%%%%%%%%%%%%%%%%%%%%%%%%%%%%%%%%%%%%%%%%%%%%%%%%%%%%%%%%%%%%%%%%%%%%%%%%  
There exists the well-known free field realization in terms of a free boson 
 and $\beta$-$\gamma$ system i.e. the Wakimoto representation \cite{Waki1986}. 
However, we introduce another realization of 
 the $\widehat{sl}_k(2)$ current algebra in terms of the bosons obtained in the previous section 
because it is useful in order to consider the insertion of the $\CK$ operator 
 which we will see in the section 5. 

The following additional boson are required:  
\be
 \phi_0(z) = - \sum_{n \in \textbf{Z}} \frac{\phi^{(0)}_{2n+1}}{2n+1} z^{-(2n+1)}, 
\ee
with 
\be
 [\phi^{(0)}_{2n+1} , \phi^{(0)}_{-2m-1}] =  k(2n+1) \delta_{n,m}. 
\ee
This has the same algebraic structure as $\phi^{(2)}(z)$ (times $\im$). 
Following \cite{HJKO0005O}, let us introduce 
\begin{align}
 \beta(z) &=  \partial \phi_0(z), \\
 x(z) &=  :(\partial\phi_1(z) + \partial\phi_2(z)) 
 \ex^{\frac{2}{k}\varphi(z)}:,  
\end{align}
where $\varphi(z) = \phi_0(z) + \phi_2(z)$ %and which satisfy 
%\begin{align}
% \beta(z_1) \beta(z_2) =& \frac{k}{2} 
% \left( \frac{1}{(z_1-z_2)^2} + \frac{1}{(z_1+z_2)^2}
% \right) + :\beta(z_1)\beta(z_2):, \\
% \beta(z_1) x(z_2) =&  
% \left(
% \frac{1}{z_1-z_2} - \frac{1}{z_1+z_2}
% \right) x(z_2) + :\beta(z_1) x(z_2):, \\
% x(z_1) x(z_2) =&  -k\frac{1}{(z_1+z_2)^2} 
% :\ex^{\frac{2}{k}\varphi(z_1)} 
% \ex^{\frac{2}{k}\varphi(z_2)}: \cr 
% &- \left( \frac{1}{z_1-z_2} - \frac{1}{z_1+z_2} \right) 
% :\ex^{\frac{2}{k}\varphi(z_1)} x(z_2): \cr 
% &+ \left( \frac{1}{z_1-z_2} + \frac{1}{z_1+z_2} \right) 
% :x(z_1) \ex^{\frac{2}{k}\varphi(z_2)} : 
% + :x(z_1) x(z_2):.
  %-\frac{k}{(z_1+z_2)^2} - \frac{2}{z_1+z_2} \beta(z_2). 
%\end{align}
and 
the symbol $::$ stands for the normal ordering defined in the standard way.
Using $\beta(z)$ and $x(z)$, 
 we can explicitly construct the $\widehat{sl}(2)$ currents of the level $k$ as follows: 
\begin{align} 
 &E(w) = \left.\frac{1}{2} (\beta(z) - x_e(z))\right|_{w=z^2}, \\
 &F(w) = \left.\frac{1}{2z^2} (\beta(z) + x_e(z))\right|_{w=z^2}, \\
 &H(w) = \left.\left(\frac{x_o(z)}{z} + \frac{k}{2z^2}\right)\right|_{w=z^2}, 
\end{align}
where 
\be
 x_e(z) = \frac{1}{2}(x(z) + x(-z)), ~~~~~
 x_o(z) = \frac{1}{2}(x(z) - x(-z)). 
\ee
Note that the currents are defined on $w$-plane. 
In fact, one can easily check 
 that $E(w)$, $F(w)$ and $H(w)$ serve precisely as the affine $\widehat{sl}(2)_k$ currents,  
\be 
\begin{split}
 &H(w_1)H(w_2) \sim \frac{2k}{(w_1 - w_2)^2}, \\
 &H(w_1)E(w_2) \sim \frac{2}{w_1 - w_2} E(w_2), \\
 &H(w_1)F(w_2) \sim \frac{-2}{w_1 - w_2} F(w_2), \\
 &E(w_1)F(w_2) \sim \frac{k}{(w_1 - w_2)^2} + \frac{1}{w_1 - w_2} H(w_2). 
\end{split}
\ee 
%from which we read off
%\begin{align}
% \phi_1(z_1) \phi_1(z_2) &\sim \frac{k+2}{2} \log (z_1^2 - z_2^2), \\
% \phi_2(z_1) \phi_2(z_2) &\sim -\frac{k}{2} \log \frac{z_1-z_2}{z_1+z_2}. 
%\end{align}
The stress tensor with the central charge $c = \frac{3k}{k+2}$ 
 can also be constructed by the Sugawara construction,  
\begin{align}
 T(w) = \frac{1}{4z^2} &\left\{ 
        \frac{1}{k} (\partial \phi_0)^2 + \frac{1}{\kappa}(\partial \phi_1)^2 
        -\frac{1}{k} (\partial \phi_2)^2 
        - \frac{1}{\kappa} \left(\partial + \frac{1}{z} \right) \partial \phi_1 \right. \notag \\ 
        &~~~~ \left.\left. + \frac{1}{z}
        \biggl(\partial \phi_1 \cosh \frac{2}{k} \varphi
        + \partial \phi_2 \sinh \frac{2}{k} \varphi \biggl) 
        + \frac{k(k+4)}{4\kappa z^2}
        \right\} \right|_{w = z^2},  
        \label{stress}
\end{align} 
where $\kappa = k+2$ as usual.

%%%%%%%%%%%%%%%%%%%%%%%%%%%%%%%%%%%%%%%%%%%%%%%%%%%%%%%%%%%%%%%%%%%%%%%%%%%%%%%%%
\subsection{spin $\bm{j/2}$ representation}
%%%%%%%%%%%%%%%%%%%%%%%%%%%%%%%%%%%%%%%%%%%%%%%%%%%%%%%%%%%%%%%%%%%%%%%%%%%%%%%%%
The operators corresponding to 
 the highest (respectively, lowest) weight state of the spin 1/2 representation are given by  
\begin{align}
 V_{1/2,1/2}(w) &= \left. \frac{1}{2} :\left( 
 \ex^{\alpha^i \phi_i(z)} + \ex^{{\alpha}^i \phi_i(-z)} 
 \right): \right|_{w=z^2} 
 = z^{\frac{k}{2\kappa}} \left. 
 :\ex^{\frac{1}{\kappa} \phi_1} 
 \cosh\left(\frac{1}{k} \varphi(z) \right): \right|_{w=z^2}, \\
 {V}_{1/2,-1/2}(w) &= \left. \frac{1}{2z} :\left( 
 \ex^{\alpha^i \phi_i(z)} - \ex^{{\alpha}^i \phi_i(-z)} 
 \right): \right|_{w=z^2} 
 = z^{\frac{k}{2\kappa}} \left.  :\ex^{\frac{1}{\kappa}  \phi_1} 
 \frac{\sinh\left(\frac{1}{k} \varphi(z) \right)}{z}: \right|_{w=z^2},   
\end{align}
where the repeated indices $i$ are summed over for $i=0,1,2$ and 
\be 
(\alpha^0 , \alpha^1, \alpha^2) 
 %= ( \widetilde{\alpha}_0 ,  -\widetilde{\alpha}_1,  -\widetilde{\alpha}_2)
 = \left( \frac{1}{k}, \frac{1}{\kappa}, \frac{1}{k}\right). 
\ee
In general, the  operators corresponding to the states belonging to 
  the spin $j/2$ representation ($j \in \bf{Z}_{\geq 0}$) are given by  
\be \label{vertex} 
 V_{j/2,j/2-m}(w) 
 = z^{\frac{jk}{2\kappa}-m} \left. :\ex^{j \alpha^1 \phi_1} 
 \cosh^{j-m}\left(\frac{1}{k} \varphi(z)\right) 
 \sinh^{m}\left(\frac{1}{k} \varphi(z)\right) :
 \right|_{w=z^2}, ~~~~~
 0 \leq m \leq j.  
\ee
In particular, the operator corresponding to the highest weight state is 
\be
 V_{j/2}(w) \left. \equiv V_{j/2,j/2}(w) %= :V^j(w): 
 = z^{\frac{kj}{2\kappa}} :\ex^{j \alpha^1 \phi_1} 
 \cosh^{j} \left(\frac{1}{k}\varphi(z) \right):
 \right|_{w=z^2}. 
\ee
The vertex operator $V_{j/2,j/2-m}(w)$ has the expected behavior,  
\be
\begin{split}
 H(w_1) V_{j/2,j/2-m}(w_2) 
 &\sim \frac{2\left(\frac{j}{2}-m\right)}{w_1-w_2} V_{j/2,j/2-m}(w_2), \\
 E(w_1) V_{j/2,j/2-m}(w_2) 
 &\sim \frac{m}{w_1-w_2} V_{j/2,j/2-m+1}(w_2), \\
 F(w_1) V_{j/2,j/2-m}(w_2) 
 &\sim \frac{j-m}{w_1-w_2} V_{j/2,j/2-m-1}(w_2), 
\end{split}
\ee
On the other hand, it is easy to check that $V_{j/2,j/2-m}(w)$ is also primary operator 
 with the scaling dimension $\Delta_j = \frac{j(j+2)}{4\kappa}$.
%\be
% T(w_1) V_{j/2,j/2-m}(w_2) 
% \sim \frac{\Delta_j}{(w_1-w_2)^2} V_{j/2,j/2-m}(w_2)  
% + \frac{1}{w_1-w_2} \partial_{w_2} V_{j/2,j/2-m}(w_2). 
%\ee
%The scaling dimension is 
%\be 
% \Delta_j = \frac{j(j+2)}{4\kappa}. 
%\ee

%%%%%%%%%%%%%%%%%%%%%%%%%%%%%%%%%%%%%%%%%%%%%%%%%%%%%%%%%%%%%%%%%%%%%%%%%%%%%%%%%
%%%%%%%%%%%%%%%%%%%%%%%%%%%%%%%%%%%%%%%%%%%%%%%%%%%%%%%%%%%%%%%%%%%%%%%%%%%%%%%%%
\section{Solutions of KZ equation}
%%%%%%%%%%%%%%%%%%%%%%%%%%%%%%%%%%%%%%%%%%%%%%%%%%%%%%%%%%%%%%%%%%%%%%%%%%%%%%%%%
Let $V=V_{j_1} \otimes V_{j_2} \otimes \cdots \otimes V_{j_N}$.  
Here $V_{j_n}, 0 \leq n \leq N$ is the $(j_n+1)$-dimensional
 lowest weight module for $\mathfrak{sl}_2$. 
The standard Chevalley basis of $\mathfrak{sl}_2$ is denoted by $\{e,f,h\}$ 
 with $[e,f]=h, [h,e]=2e, [h,f]=-2f$.  
The lowest weight state $v_{j} \in V_j$ is defined by
\be
  f \cdot v_{j} = 0, ~~~~ h \cdot v_{j} = -j v_{j}, ~~~~
  e^{j+1} \cdot v_{j} = 0. 
\ee
The KZ equation \cite{KZ} (for review, see \cite{EFK1997L}) is written as 
\be
 \kappa \frac{\partial}{\partial w_n} \Psi(\bm{w}) 
  = \left( \sum_{m=1, m \neq n}^{N} \frac{\Omega_{mn}}{w_m - w_n} \right) 
  \Psi(\bm{w}), 
  ~~~~~~~ n = 1, \cdots, N.  
 \label{KZ}
\ee 
where $\Psi(\bm{w})$ is the function which takes value in $V$ and    
\be
  \Omega_{mn}= e_m f_n + f_m e_m + \frac{1}{2} h_m h_n. 
\ee
Here $x_n$ stands for the action of $x \in \mathfrak{sl}_2$ on $V_{j_n}$, i.e. 
\begin{align}
  x_n &= 1 \otimes 1 \otimes \cdots \otimes \stackrel{\substack{n\\ \vee}}{x} 
  \otimes \cdots \otimes 1,  \\
  x_m y_n &= 
  1 \otimes 1 \otimes \cdots \otimes \stackrel{\substack{m\\ \vee}}{x} \otimes 
  \cdots \otimes \stackrel{\substack{n\\ \vee}}{y} \otimes \cdots \otimes 1
\end{align}
In this section, 
 we want to get the integral formulas of the solutions of \eqref{KZ}
 in the free field realization constructed in the previous section.

%%%%%%%%%%%%%%%%%%%%%%%%%%%%%%%%%%%%%%%%%%%%%%%%%%%%%%%%%%%%%%%%%%%%%%%%%%%%%%%%%
\subsection{A simple solution}
%%%%%%%%%%%%%%%%%%%%%%%%%%%%%%%%%%%%%%%%%%%%%%%%%%%%%%%%%%%%%%%%%%%%%%%%%%%%%%%%%
Let us define 
\begin{gather}
 X(z) = :\cosh\left( \frac{1}{k} (\phi_0 + \phi_2) (z)\right):, ~~~~~
 Y(z) = \frac{1}{z} :\sinh\left( \frac{1}{k} (\phi_0 + \phi_2) (z)\right):, \\
 Z_{j}(z) =  z^{\frac{kj}{2\kappa}}  :\ex^{\frac{j}{\kappa} \phi_1(z)}:.  
\end{gather}
Then the operator \eqref{vertex} is expressed by
\be
 V_{j/2,j/2-m}(w) = Z_j(z) X(z)^{j-m} Y(z)^m \biggl|_{w=z^2}. 
\ee
Note that 
\begin{gather} 
 X(z_1)X(z_2) = :X(z_1)X(z_2):, ~
 Y(z_1)Y(z_2) = :Y(z_1)Y(z_2):, ~
 X(z_1)Y(z_2) = :X(z_1)Y(z_2):,   \\
 \prod_{i=1}^N Z_{j_i} (z_i) 
 = \prod_{m<n}^N (z_m^2 - z_n^2)^{\frac{j_mj_n}{2\kappa}} 
 {\textbf{:}}\prod_{n=1}^N Z_{j_n} (z_n){\textbf :}. 
\end{gather}
Let us choose the Fock vacuum $\ket{\Omega}$ 
 and the conjugate $\bra{\Omega}$ as  
\be
\begin{split}
 &\phi^{(0)}_{2n+1} \ket{\Omega} = 0, ~~~
 \phi^{(1)}_{2n} \ket{\Omega} = 0, ~~~
 \phi^{(2)}_{2n+1} \ket{\Omega} =0, ~
 n \geq 0, \\
 &\bra{\Omega} \phi^{(0)}_{2n+1}  = 0, ~~~
 \bra{\Omega} \phi^{(1)}_{2n} = 0, ~~~
 \bra{\Omega} \phi^{(2)}_{2n+1} = 0, ~~~
 n < 0, 
\end{split}
\ee 
The highest weight state of the spin $j/2$ representation is given by 
\be
 \ket{j} %= \lim_{z \to 0} V_{j/2}(z) \ket{\Omega} 
  = \ex^{\frac{1}{\kappa}\left(j - \frac{k}{2} \right)Q} \ket{\Omega} , ~~~~~~ 
 \bra{j} = \bra{\Omega} \ex^{-\frac{1}{\kappa}\left(j - \frac{k}{2}\right)Q}.  
\ee
We denote  by $\CH_j$ the highest weight module over $\widehat{sl}(2)_k$ 
 generated from $\ket{j}$. 
Then the operator $V_{j/2,j/2-m}(w)$ plays a role to map $\CH_{j_0}$ onto 
 $\CH_{j_0 + j}$ for any $j_0 \in \bf{Z}_{\geq 0}$.
From the product of $N$ $V_{j/2}$'s we obtain  
\be
 \bra{j} \prod_{n=1}^N V_{j_n/2}(w_n) \ket{0} 
 = \prod_{m<n}^N (w_m - w_n)^{\frac{j_mj_n}{2\kappa}}
 \equiv \psi_0(\bm{w}), 
 ~~~~~~ j = \sum_{n=1}^N j_n. 
\ee
Consequently the simple solution of \eqref{KZ} is given by 
\be
 \Psi_0(\bm{w}) = \psi_0(\bm{w}) v 
 = \prod_{m<n}^N (w_m - w_n)^{\frac{j_mj_n}{2\kappa}} v 
 = \bra{j} \prod_{n=1}^N V_{j_n/2}(z_n) \ket{0} v, ~~~~~~
 j = \sum_{n=1}^N j_n,  
\ee
where  $v = v_{j_1} \otimes v_{j_2} \otimes \cdots \otimes v_{j_N} \in V$.

%%%%%%%%%%%%%%%%%%%%%%%%%%%%%%%%%%%%%%%%%%%%%%%%%%%%%%%%%%%%%%%%%%%%%%%%%%%%%%%%%
\subsection{screening charge}
%%%%%%%%%%%%%%%%%%%%%%%%%%%%%%%%%%%%%%%%%%%%%%%%%%%%%%%%%%%%%%%%%%%%%%%%%%%%%%%%%
The screening current is defined by \footnote{The screening current 
 \eqref{screening} can be obtained by taking the root of unity limit
 of the $q$-deformed screening current $\widetilde{S}(z) = \ex^{\widetilde{\varphi}(z)}$ 
 up to the overall factor. 
} 
\be 
  %S(z) :=  -\frac{2}{k+2} z^{\frac{2}{k+2}} \partial \phi_2(z) 
  %       \ex^{-\frac{2}{k+2} \phi_1(z)},  
  S(\tau) =  %- 
  t^{-\frac{k}{\kappa}} \partial \phi_2(t) 
         \ex^{-\frac{2}{\kappa} \phi_1(t)} \biggl|_{\tau = t^2}, 
  \label{screening} 
\ee
which satisfies 
\be
\begin{split}
 \beta(z) S(\tau) &\sim 0, \\
 x_e(z) S(\tau) &\sim \frac{\partial}{\partial \tau}
 A_e(z,\tau), \\
 \frac{x_o(z)}{z} S(\tau) &\sim \frac{\partial}{\partial \tau}
 A_o(z,\tau), 
\end{split}
\ee
where 
\begin{align}
 A_e(z,\tau) &= \kappa \frac{t^{\frac{k+4}{k+2}}}{z^2 - t^2} \left(
  \ex^{\frac{2}{k}(\phi_0 + \phi_2)(t)} + \ex^{-\frac{2}{k}(\phi_0 + \phi_2)(t)}
  \right) \ex^{-\frac{2}{k+2} \phi_1(t)} \biggl|_{\tau = t^2}, \\
 A_o(z,\tau) &= \kappa \frac{t^{\frac{k+4}{k+2}}}{z^2 - t^2} \frac{1}{t}\left(
  \ex^{\frac{2}{k}(\phi_0 + \phi_2)(t)} - \ex^{-\frac{2}{k}(\phi_0 + \phi_2)(t)}
  \right) \ex^{-\frac{2}{k+2} \phi_1(t)} \biggl|_{\tau = t^2}. 
\end{align} 
The screening charge defined by 
\be
 U = \int_{C} \de\tau S(\tau), ~~~~~
 %Q_2 = \oint \de z \eta(z)
\ee
commutes with the $\widehat{sl}(2)_k$ currents $E(w), H(w), F(w)$. 
Here we postulate the cycle $C$ on $w$-plane is chosen appropriately.

%%%%%%%%%%%%%%%%%%%%%%%%%%%%%%%%%%%%%%%%%%%%%%%%%%%%%%%%%%%%%%%%%%%%%%%%%%%%%%%%%
\subsection{intertwining operator}
%%%%%%%%%%%%%%%%%%%%%%%%%%%%%%%%%%%%%%%%%%%%%%%%%%%%%%%%%%%%%%%%%%%%%%%%%%%%%%%%%
In this section, we consider the intertwining operator 
 $\Phi_j^m(z) : \CH_{j_0} \to \CH_{j_0+j-2m} \otimes V_j$, 
 $^{\forall} j_0 \in \textbf{Z}_{\geq 0}$.  
Let us introduce the following formal operator:   
\be
 \gamma(z) =  X(z)^{-1}Y(z). 
\ee
The intertwining operator of level $m = 0$ can be constructed in terms of 
 $Z_j(z)$, $X(z)$ and $Y(z)$ as   
\be
 \Phi_j^0(z) u = Z_j(z) X^j(z) \ex^{-\gamma(z) \otimes e} (u \otimes v_j), 
\ee
where $u \in \CH_{j_0}$ and $v_j \in V_j$ is the lowest weight state. 
In fact, $\Phi_j^0(z)$ satisfies the intertwining relation, 
\be
 %[J_n^A \otimes 1,  \Phi_j^0(z)] = (z^2)^n (1 \otimes (-A)) \Phi_j^0(z), 
 \Phi_j^0(z) J_n^A = (J_n^A \otimes 1 + (z^2)^n \cdot 1 \otimes A)
  \Phi_j^0(z)
 ~~~~~ A= e,f,h.   
\ee
By using the intertwining operators, $\Psi_0(\bm{w})$ is given by  
\be
 \Psi_0(\bm{w}) = \bra{j} \prod_{n=1}^N \Phi^0_{j_n}(z_n) \ket{0}, 
 ~~~~~~ j = \sum_{n=1}^N j_n.
\ee 
The general $m$ intertwining operator can be obtained by 
 multiplying $m$ screening charges to $\Phi_j^0(z)$,    
\be
  \Phi_j^m(z) u = \Phi_j^0(z) U^m, 
\ee
and the solutions of the KZ equation are given by 
\be
 \Psi_m(\bm{w}) = \bra{j-2m} \prod_{i=1}^N \Phi_{j_i}^{m_i} (z_i) \ket{0}, 
 ~~~~~  j = \sum_{i=1}^N j_i , ~~ m = \sum_{i=1}^N m_i. 
\ee 
which reproduces the well-known integral formulas. 
For example, we obtain, in the case of $m=1$,  
\begin{align}
 \Psi_1(\bm{w}) 
 &= \bra{j-2} \prod_{n=1}^{N-1} \Phi_{j_n}^0(z_n) \Phi_{j_N}^1(z_N) \ket{0} \notag\\
 &= \prod_{m<n}^N  (w_m - w_n)^{\frac{j_mj_n}{2\kappa}} 
 \int_C \de \tau 
 %\tau^{-\frac{k}{2\kappa}} 
 \prod_{n=1}^N (w_n - \tau)^{-\frac{j_n}{\kappa}} 
 \left( \sum_{n=1}^N \frac{- e_n}{w_n-\tau} \right) v, 
\end{align}
and in the case of $m=2$, 
\begin{align}
 \Psi_2(\bm{w}) 
 &= \bra{j-4} \prod_{n=1}^{N-1} \Phi_{j_n}^{0}(z_n) \Phi_{j_N}^{2}(z_N) \ket{0} \notag\\
 &= \prod_{m<n}^N  (w_m - w_n)^{\frac{j_mj_n}{2\kappa}} 
 \int_{C} \de \bm{\tau}  
 %(\tau_1\tau_2)^{-\frac{k}{2\kappa}} 
 \prod_{n=1}^N \prod_{p=1}^2  (w_n - \tau_p)^{-\frac{j_n}{\kappa}} 
 (\tau_1 - \tau_2)^{\frac{2}{\kappa}}
 \notag \\
 & \hspace{7cm} 
 \times \left( \sum_{m=1}^N \sum_{n=1}^N \frac{e_m e_n}{(w_m-\tau_1)(w_n - \tau_2)}
  \right) v, 
\end{align}
where $\bm{\tau} = (\tau_1, \tau_2)$ and $C = C_1 \times C_2$. 
Here we have used 
\begin{align}
 \Phi_j^0(z) S(\tau) 
 &= (w-\tau)^{-\frac{j}{\kappa}} \left\{
  :\Phi_j^0(z) S(\tau): 
  -\frac{\tau^{-\frac{k}{2\kappa}}}{w-\tau} \bigl(
  (1 \otimes e) -(w \otimes f) \bigl) 
  :\ex^{-\frac{2}{\kappa} \phi_1(t)} \Phi_j^0(z): 
 \right\}, \\
 S(\tau)S(\tau') 
 &=  (\tau - \tau')^{\frac{2}{\kappa}} 
 \left\{ :S(\tau)S(\tau'):
  -\frac{k(\tau\tau')^{- \frac{k}{2\kappa}}(\tau + \tau')}{(\tau - \tau')^2} 
  :\ex^{-\frac{2}{\kappa} (\phi_1(t) + \phi_1(t'))}: 
 \right\}. 
\end{align}

%%%%%%%%%%%%%%%%%%%%%%%%%%%%%%%%%%%%%%%%%%%%%%%%%%%%%%%%%%%%%%%%%%%%%%%%%%%%%%%%%
%%%%%%%%%%%%%%%%%%%%%%%%%%%%%%%%%%%%%%%%%%%%%%%%%%%%%%%%%%%%%%%%%%%%%%%%%%%%%%%%%
\section{Insertion of $\CK$ operator}
%%%%%%%%%%%%%%%%%%%%%%%%%%%%%%%%%%%%%%%%%%%%%%%%%%%%%%%%%%%%%%%%%%%%%%%%%%%%%%%%%
%\subsection{four-point current block}
%%%%%%%%%%%%%%%%%%%%%%%%%%%%%%%%%%%%%%%%%%%%%%%%%%%%%%%%%%%%%%%%%%%%%%%%%%%%%%%%%
In the paper \cite{AT1005}, the authors proposed 
 the insertion of the $\CK$ operator to the $\widehat{sl}(2)_k$ current block current block 
 in order to establish the AGT relation in the presence of a full surface operator.
The generalization to the $\widehat{sl}(n)_k$ current block is also discussed in \cite{KPPW1008}.
The $\CK$ operator is defined by 
\footnote{Since the $\CK$ operator we will use below   
 is equal to $\CK^{\dag}(1,1)$ presented in \cite{KPPW1008}, 
 we denote it by $\CK^{\dag}$.} 
\be
 \CK^{\dag} = \exp \left( \sum_{n=1}^{\infty} \frac{1}{2n-1}
  [J^+_{n-1} + J^-_{n}] \right). 
\ee
In our formalism, the $\CK$ operator
 has the following  simple expression in terms of $\phi_0(z)$: 
 \be
 %\CK = \ex^{-\phi^{(-)}_0(1)}, ~~~~~~
 \CK^{\dag} = \ex^{-\phi^{(+)}_0(1)}, 
 \label{K-op:boson}
\ee
where $\phi^{(+)}_0(z)$ is the positive modes of $\phi_0(z)$.\footnote{The original 
 $\CK$ operator in \cite{AT1005} can be realized 
  by the negative modes of $\phi_0$.}
Therefore it is meaningful to consider the four-point correlation function  
 with a $\CK$ operator insertion,  
\begin{align}
 \widetilde{\Psi}_{m_1,m_2}(w_2) 
 &\equiv \bra{j-2m} {\Phi}_{j_1}^{m_1}(1) \ex^{-\phi^{(+)}_0(1)} 
 \Phi_{j_2}^{m_2}(z_2) \ket{j_3}. % \otimes v_{j_3}. 
 %= \lim_{\substack{z_1 \to 1 \\z_3 \to 0}}
 % \bra{j-2m} {\Xi}_{j_1}^{m_1}(z_1) \Phi_{j_2}^{m_2}(z_2) 
 % V_{j_3}(z_3) \ket{0},   
\end{align}
Since the screening current $S(z)$ does not include $\phi_0(z)$, 
 it is trivial that the $\CK$ operator commutes with $U$. 
However, the action of $\ex^{-\phi^{(+)}_0(1)}$ on  ${\Phi}_{j}^{0}(z)$ 
 yields nontrivial result and we obtain 
\be
 \widetilde{\Psi}_{m_1,m_2}(w_2) = \bra{j-2m} {\Phi}_{j_1}^{m_1}(1) 
 \widetilde{\Phi}_{j_2}^{m_2}(z_2) \ket{j_3}, % \otimes v_{j_3},
\ee
where 
\begin{align}
 \widetilde{\Phi}_{j}^{m}(z) u 
 &= \widetilde{\Phi}_{j}^{0}(z) U^m  u \cr 
 &= Z_j(z) \widetilde{X}(z) \ex^{-\widetilde{\gamma}(z) \otimes e} 
 U^m u \otimes v_j,  
\end{align}
with 
\be 
\begin{split}
 \widetilde{X}(z) &= X(z) + z^2 Y(z), \\
 \widetilde{Y}(z) &= Y(z) + X(z), \\
 \widetilde{\gamma}(z) &= \widetilde{X}^{-1}(z) \widetilde{Y}(z). 
\end{split}
\ee  
It is easy to get the explicit 
  integral formulas for the four point current block with the $\CK$ operator insertion. 
For example, 
\begin{align}
 \widetilde{\Psi}_{0,0}(w_2)
 %&=\bra{j} \Xi_{j_1}^0(1) \Phi_{j_2}^0(z_2) \ket{j_3} \notag\\ 
 %&= \lim_{\substack{z_1 \to 1 \\ z_3 \to 0}} \widetilde{\Psi}_0(\bm{w}) \notag \\
 &= (1 - w_2)^{\frac{j_1j_2}{2\kappa}-\frac{j_2}{2}} w_2^{\frac{j_2j_3}{2\kappa}} 
    \left( v_{j_1} \otimes \ex^{-e} v_{j_2} % \otimes v_{j_3} 
    \right),\label{solution0}\\
 \widetilde{\Psi}_{0,1}(w_2) 
 %&= \bra{j-2} \Xi_{j_1}^0(1) \Phi_{j_2}^0(z_2) Q \ket{j_3} \notag\\ 
 &= (1 - w_2)^{\frac{j_1j_2}{2\kappa}-\frac{j_2}{2}} w^{\frac{j_2j_3}{2\kappa}}
 \int_C \de \tau (1-\tau)^{-\frac{j_1}{\kappa}} (w_2-\tau)^{-\frac{j_2}{\kappa}}
 \tau^{-\frac{j_3}{\kappa}} \notag\\    %\left(j_3 +\frac{k}{2}\right)} \notag\\ 
 & \hspace{6cm} \left\{ \frac{-e_1}{1 - \tau} + \frac{-e_2 + w_2 f_2}{w_2 - \tau} 
  \right\} 
  \left(v_{j_1} \otimes \ex^{-e} v_{j_2} %\otimes v_{j_3}
  \right).\label{solution1}
\end{align}
%%%%%%%%%%%%%%%%%%%%%%%%%%%%%%%%%%%%%%%%%%%%%%%%%%%%%%%%%%%%%%%%%%%%%%%%%%%%%%%%%
%%%%%%%%%%%%%%%%%%%%%%%%%%%%%%%%%%%%%%%%%%%%%%%%%%%%%%%%%%%%%%%%%%%%%%%%%%%%%%%%
\section{Summary}
%%%%%%%%%%%%%%%%%%%%%%%%%%%%%%%%%%%%%%%%%%%%%%%%%%%%%%%%%%%%%%%%%%%%%%%%%%%%%%%%%
To summarize, we have reproduced the free field realization of 
 the $sl(2)_k$ current algebra and the integral formulas of the KZ equation
 by using three types of chiral bosons which are obtained from the $q$-deformed boson 
 in the root of unity limit. 
In addition, we have derived the integral formulas for the modified four point  current blocks. 

%What kind of equation do the modified blocks satisfy? 
Finally, we give a comment on the modification of KZ equation.
The insertion of the $\CK$ operator would modify the original KZ equation. 
Now, let us examine the OPE with the stress-energy tensor \eqref{stress}, 
\begin{align}
 T(w_1) \widetilde{\Phi}_{j}^{0}(w_2) 
 \sim \frac{1}{(w_1-w_2)^2} \frac{j(j+2)}{4 \kappa} \widetilde{\Phi}_{j}^{0}(w_2) 
 + \frac{1}{w_1-w_2} \partial_{w_2}\widetilde{\Phi}_{j}^{0}(w_2)  \cr
 - \frac{1}{w_1-w_2} \left(
 j \widetilde{X}^{-1} Y \otimes 1 + \widetilde{X}^{-2} \widetilde{Y} Y \otimes e
 \right) \widetilde{\Phi}_{j}^{0}(w_2).  
\end{align}
The last term is the additional one. 
The OPEs with $H(w)$, $F(w)$ and $E(w)$ have also the extra terms.  
The modification of the KZ equation  is provided 
 by these unusual behaviors in $\widetilde{\Phi}_{j}^{0}(z)$ with respect to the Virasoro algebra
 and the current algebra.  
Such an equation may have the solution displayed in this paper, 
 for example, \eqref{solution0} and \eqref{solution1} and 
 may be related to the quantum isomonodromy equation proposed in \cite{Yama1011}.

%Moreover, we have considered the $sl(2)_k$ current algebra 
% and the the modification of the four point current blocks in this paper.
%The extension to the $sl(n)_k$ current algebra and 
% the general $N$-point blocks should also to be constructed. 

%%%%%%%%%%%%%%%%  acknowledgements %%%%%%%%%%%%%%%%%%%%%%
\section*{Acknowledgments}
We thank Hiroshi Itoyama, Takeshi Oota, Sho Deguchi and Hiroaki Kanno for valuable discussions. 
This research is supported in part by the Grant-in-Aid 
for Scientific Research from the Ministry of Education, Science and Culture, Japan (15K05059).
Support from JSPS/RFBR bilateral collaboration 
``Faces of matrix models in quantum field theory and statistical mechanics'' 
 is gratefully appreciated.
%%%%%%%%%%%%%%%%%%%%%%%%%%%%%%%%%%%%%%%%%%%%%%%%%%%%%%%%%%%

%%%%%%%%%%%%%%%%%%%%%%%%%%%%%%%%%%%%%%%%%%%%%%%%%%%%%%%%%%%%%%%%%%%%%%%%%%%%%%%%
%%%%%%%%%%%%%%%%%%%%%%%%%%%%%%%%%%%%%%%%%%%%%%%%%%%%%%%%%%%%%%%%%%%%%%%%%%%%%%%%
\bibliographystyle{ytphys}

\begin{thebibliography}{99}

\bibitem{AGT}
L.~F.~Alday, D.~Gaiotto and Y.~Tachikawa, 
  ``Liouville Correlation Functions from Four-dimensional Gauge Theories,'' 
  Lett.\ Math.\ Phys.\ {\bf 9}, 167-197 (2010) [arXiv:0906.3219 [hep-th]].
  
\bibitem{Nekrasov}
N. Nekrasov, ``Seiberg-Witten prepotential from instanton counting," 
Adv. Theor. Math. Phys. {\bf 7}, 831-864 (2004) [arXiv:hep-th/0206161].

\bibitem{BPZ}
A. A. Belavin, A. M. Polyakov, and A. B. Zamolodchikov, 
``Infinite conformal symmetry in two-dimensional quantum field theory," 
Nucl. Phys. B {\bf 241}, 333-380 (1984).

\bibitem{Wyl0907}
N.~Wyllard, ``$A_{N-1}$ conformal Toda field theory correlation functions from
  conformal $\mathcal{N}=2$ $SU(N)$ quiver gauge theories,'' 
  JHEP {\bf 0911}, 002 (2009) [arXiv:0907.2189 [hep-th]].
  
\bibitem{MM0908}
A. Mironov and A. Morozov, 
``On AGT relation in the case of U(3)," 
Nucl. Phys. B {\bf 825}, 1-37 (2010)
[arXiv:0908.2569 [hep-th]]. 

\bibitem{HKS1012}
L. Hollands, C. A. Keller, J. Song,
``From SO/Sp instantons to W-algebra blocks,"  
JHEP {\bf 1103} (2011) 053 [arXiv:1012.4468 [hep-th]]. 

\bibitem{HKS1107}
L. Hollands, C. A. Keller, J. Song,
``Towards a 4d/2d correspondence for Sicilian quivers," 
JHEP 1110 (2011) 100 
[arXiv:1107.0973 [hep-th]]. 

\bibitem{KMST1111}
The ABCDEFG of Instantons and W-algebras 
C. A. Keller, N. Mekareeya, J. Song and Y. Tachikawa,
JHEP 1203 (2012) 045 
[arXiv:1111.5624 [hep-th]]. 


\bibitem{DV}
R.~Dijkgraaf and C.~Vafa, ``Toda Theories, Matrix Models, Topological Strings,
  and $N = 2$ Gauge Systems,'' [arXiv:0909.2453 [hep-th]].

\bibitem{IMO}
H.~Itoyama, K.~Maruyoshi and T.~Oota,  ``The Quiver Matrix Model and 
  2d-4d Conformal Connection,'' Prog.\ Theor.\ Phys.\ {\bf 123}, 957-987 (2010)
  [arXiv:0911.4244 [hep-th]].

\bibitem{MMS0911}
A.~Mironov, A.~Morozov and Sh.~Shakirov, ``Matrix Model Conjecture for 
  Exact BS Periods and Nekrasov Functions,'' 
  JHEP {\bf 1002}, 030 (2010) [arXiv:0911.5721 [hep-th]].

\bibitem{FL0912}
V. A. Fateev and A. V. Litvinov,
``On AGT conjecture," 
JHEP 1002 (2010) 014 [arXiv:0912.0504 [hep-th]].

\bibitem{MMS1001}
A.~Mironov, A.~Morozov and Sh.~Shakirov,  ``Conformal blocks as 
  Dotsenko-Fateev Integral Discriminants,'' 
  J.\ Mod.\ Phys.\ A {\bf 25}, 3173-3207 (2010) [arXiv:1001.0563 [hep-th]].

\bibitem{IO5}
H.~Itoyama and T.~Oota, ``Method of generating $q$-expansion coefficients for
  conformal block and $\mathcal{N}=2$ Nekrasov function by $\beta$-deformed
  matrix model,'' Nucl.\ Phys.\ B {\bf 838}, 298-330 (2010) 
  [arXiv:1003.2929 [hep-th]].

\bibitem{MMM1003}
A.~Mironov, A.~Morozov and And.~Morozov, ``Matrix model version of AGT
  conjecture and generalized Selberg integrals,'' Nucl.\ Phys.\ B {\bf 843},
  534-557 (2011) [arXiv:1003.5752 [hep-th]].

\bibitem{IOYone}
H.~Itoyama, T.~Oota and N.~Yonezawa, ``Massive scaling limit of the 
  $\beta$-deformed  matrix model of Selberg type,'' 
  Phys.\ Rev.\ D {\bf 82}, 085031 (2010) [arXiv:1008.1861 [hep-th]].

\bibitem{MMS1012}
A.~Mironov, A.~Morozov and Sh.~Shakirov, ``A direct proof of AGT conjecture at
  $\beta = 1$,'' JHEP {\bf 1102}, 067 (2011) [arXiv:1012.3137 [hep-th]].

\bibitem{BMT1104}
G. Bonelli, K. Maruyoshi and A. Tanzini, 
``Quantum Hitchin Systems via beta-deformed Matrix Models,"
arXiv:1104.4016 [hep-th]. 

\bibitem{KMZ1306}
S. Kanno, Y. Matsuo and H. Zhang, 
``Extended Conformal Symmetry and Recursion Formulae for Nekrasov Partition Function,"
JHEP {\bf 1308}, 028 (2013) [arXiv:1306.1523 [hep-th]]. 

\bibitem{MS1307}
A. Morozov and A. Smirnov, 
``Towards the Proof of AGT Relations with the Help of the Generalized Jack Polynomials,"
Lett. Math. Phys. {\bf 104} 585-612 (2014) [arXiv:1307.2576 [hepth]].

\bibitem{MMZ1312}
S. Mironov, And. Morozov and Y. Zenkevich, 
``Generalized Jack polynomials and the AGT relations for the SU(3) group," 
JETP Lett. {\bf 99}, 109-113 (2014) [arXiv:1312.5732 [hep-th]].

\bibitem{MRZ1405}
Y. Matsuo, C. Rim and H. Zhang, 
``Construction of Gaiotto states with fundamental multiplets through Degenerate DAHA," 
JHEP {\bf 1409} (2014) 028 
[arXiv:1405.3141 [hep-th]]. 


\bibitem{BF1105}
V. Belavin and B. Feigin, 
``Super Liouville conformal blocks from N = 2 SU(2) quiver gauge theories," 
JHEP {\bf 1107}, 079 (2011) [arXiv:1105.5800 [hep-th]].

\bibitem{NT1106}
T. Nishioka and Y. Tachikawa, 
``Central charges of para-Liouville and Toda theories from M-5-branes," 
Phys. Rev. D {\bf 84}, 046009 (2011) [arXiv:1106.1172 [hep-th]].

\bibitem{BMT1106}
G. Bonelli, K. Maruyoshi and A. Tanzini,
``Instantons on ALE spaces and Super Liouville Conformal Field Theories," 
JHEP {\bf 1108} (2011) 056 
[arXiv:1106.2505 [hep-th]]. 

\bibitem{BMT1107}
G. Bonelli, K. Maruyoshi and A. Tanzini, 
``Gauge Theories on ALE Space and Super Liouville Correlation Functions," 
Lett. Math. Phys. {\bf 101} (2012) 103-124 
[arXiv:1107.4609 [hep-th]].  

\bibitem{BBB1106}
A. Belavin, V. Belavin and M. Bershtein, ``Instantons and 2d Superconformal field theory," 
JHEP {\bf 1109}, 117 (2011) [arXiv:1106.4001 [hep-th]]. 

\bibitem{Wyll1109}
N. Wyllard, ``Coset conformal blocks and $\mathcal{N}=2$ gauge theories," 
arXiv:1109.4264 [hep-th].  

\bibitem{Ito1110}
Y. Ito, 
``Ramond sector of super Liouville theory from instantons on an ALE space," 
Nucl.Phys. B861 (2012) 387-402 [arXiv:1110.2176 [hep-th]].

\bibitem{AT1110}
M. N. Alfimov and G. M. Tarnopolsky, 
``Parafermionic Liouville field theory and instantons on ALE spaces," 
JHEP 1202 (2012) 036 [arXiv:1110.5628 [hep-th]]. 

\bibitem{BW1205}
V. Belavin and N. Wyllard, 
``$\mathcal{N}=2$ superconformal blocks and instanton partition functions," 
JHEP {\bf 1206} (2012) 173 [arXiv:1205.3091 [hep-th]].

\bibitem{BM1210}
A. Belavin and B. Mukhametzhanov, 
``$\mathcal{N}=1$ superconformal blocks with Ramond fields from AGT correspondence,"  
JHEP 1301 (2013) 178 [arXiv:1210.7454 [hep-th]].

\bibitem{AY0910}
H. Awata and Y. Yamada, 
``Five-dimensional AGT Conjecture and the Deformed Virasoro Algebra," 
JHEP {\bf 1001} (2010) 125 [arXiv:0910.4431 [hep-th]].

\bibitem{AY1004}
H. Awata and Y. Yamada, 
  ``Five-Dimensional AGT Relation and the Deformed $\beta$-Ensemble,'' 
  Prog.\ Theor.\ Phys.\ {\bf 124}, 227-262 (2010) 
  [arXiv:1004.5122 [hep-th]]. 

\bibitem{AFHKSY1106}
H. Awata, B. Feigin, A. Hoshino, M. Kanai, J. Shiraishi and S. Yanagida, 
``Notes on Ding-Iohara algebra and AGT conjecture," 
arXiv:1106.4088 [math-ph]. 

\bibitem{MMSS1105}
A. Mironov, A. Morozov, S. Shakirov, and A. Smirnov, 
``Proving AGT conjecture as HS duality: extension to five dimensions," 
Nucl. Phys. B {\bf 855} (2012) 128-151 [arXiv:1105.0948 [hep-th]].

\bibitem{Ohku1404}
Y. Ohkubo, 
``Existence and Orthogonality of Generalized Jack Polynomials and Its $q$-Deformation," 
arXiv:1404.5401 [math-ph].

\bibitem{Zenk1412}
Y. Zenkevich, 
``Generalized Macdonald polynomials, spectral duality for conformal
blocks and AGT correspondence in five dimensions," 
JHEP {\bf 1405} (2015) 131 [arXiv:1412.8592 [hep-th]]. 

\bibitem{MZ1510}
A. Morozov and Y. Zenkevich, 
``Decomposing Nekrasov Decomposition," 
 arXiv:1510.01896 [hep-th]. 

\bibitem{IKY1511}
A. Iqbal, C. Koz\c{c}az and S.-T. Yau, 
``Elliptic Virasoro Conformal Blocks," 
arXiv:1511.00458 [hep-th].

\bibitem{Nier1511}
F. Nieri, ``An elliptic Virasoro symmetry in 6d," 
arXiv:1511.00574 [hep-th]. 

\bibitem{IOY2}
H.~Itoyama, T.~Oota and R.~Yoshioka,
  ``2d-4d Connection between $q$-Virasoro/W Block at Root of Unity Limit 
  and Instanton Partition Function on ALE Space,''
  Nucl.\ Phys.\ B {\bf 877}, 506-537 (2013) [arXiv:1308.2068 [hep-th]].

\bibitem{IOY3} 
H.~Itoyama, T.~Oota and R.~Yoshioka,
``q-Virasoro/W Algebra at Root of Unity and Parafermions," 
Nucl. Phys. B {\bf 889} 25-35 (2014) [arXiv:1408.4216 [hep-th]]. 

\bibitem{IOYconf}
H. Itoyama, T. Oota, R. Yoshioka, 
``q-Virasoro algebra at root of unity limit and 2d-4d connection," 
J. Phys. Conf. Ser. {\bf 474} (2013) 012022. 

\bibitem{BBT1211B}
A. A. Belavin, M. A. Bershtein and  G. M. Tarnopolsky,
``Bases in coset conformal field theory from AGT correspondence 
 and Macdonald polynomials at the roots of unity,"   
JHEP 1303 (2013) 019 [arXiv:1211.2788 [hep-th]].


\bibitem{AGGTV} 
L. F. Alday, D. Gaiotto, S. Gukov, Y. Tachikawa and H. Verlinde, 
``Loop and surface operators in $\mathcal{N}=2$ gauge theory and Liouville modular geometry,"  
 JHEP {\bf 1001}, 113 (2010) [arXiv:0909.0945 [hep-th]].

\bibitem{Guko1412}
S. Gukov, ``Surface Operators," 
Math. Phys. Stud. (2016) 223-259 
[arXiv:1412.7127[hep-th]].

\bibitem{KPW1004}
C. Koz\c{c}az, S. Pasquetti and N. Wyllard,
``A \& B model approaches to surface operators and Toda theories," 
JHEP {\bf 1008} (2010) 042 [arXiv:1004.2025 [hep-th]].

\bibitem{DGH1006}
T. Dimofte, S. Gukov and L. Hollands,
``Vortex Counting and Lagrangian 3-manifolds," 
Lett. Math. Phys. {\bf 98} (2011) 225-287 [arXiv:1006.0977 [hep-th]].

\bibitem{MT1006}
K. Maruyoshi and M. Taki,
``Deformed Prepotential, Quantum Integrable System and Liouville Field Theory," 
Nucl.Phys. B {\bf 841} (2010) 388-425 [arXiv:1006.4505 [hep-th]]. 

\bibitem{Taki1007}
M. Taki, ``Surface Operator, Bubbling Calabi-Yau and AGT Relation," 
JHEP 1107 (2011) 047 [arXiv:1007.2524 [hep-th]].

\bibitem{Gaio0908A}
D. Gaiotto, ``Asymptotically free $\mathcal{N}=2$ theories and irregular conformal blocks," 
J. Phys. Conf. Ser. {\bf 462} (2013) 1, 012014 [arXiv:0908.0307 [hep-th]].

\bibitem{MMM0909O}
A. Marshakov, A. Mironov and A. Morozov, 
``On non-conformal limit of the AGT relations," 
Phys. Lett. B {\bf 682}, 125 (2009) [arXiv:0909.2052 [hep-th]].

\bibitem{AFKMY1008}
H. Awata, H. Fuji, H. Kanno, M. Manabe, Y. Yamada, 
``Localization with a Surface Operator, Irregular Conformal Blocks 
  and Open Topological String," 
Adv. Theor. Math. Phys. {\bf 16} (2012) 3, 725-804   
[arXiv:1008.0574 [hep-th]].


\bibitem{AT1005}
L. F. Alday and Y. Tachikawa, 
``Affine $SL(2)$ conformal blocks from 4d gauge theories," 
Lett.Math.Phys. {\bf 94} (2010) 87-114 [arXiv:1005.4469 [hep-th]].

\bibitem{KPPW1008}
C. Koz\c{c}az, S. Pasquetti, F. Passerini and N. Wyllard, 
``Affine $sl(N)$ conformal blocks from $\mathcal{N}=2$ $SU(N)$ gauge theories" 
JHEP 1101 (2011) 045 [arXiv:1008.1412 [hep-th]].

\bibitem{Wyll1011}
N. Wyllard, 
``W-algebras and surface operators in $\mathcal{N}=2$ gauge theories," 
J.Phys. A44 (2011) 155401 [arXiv:1011.0289 [hep-th]]. 

\bibitem{Wyll1012}
N. Wyllard, 
``Instanton partition functions in $\mathcal{N}=2$ SU(N) gauge theories 
 with a general surface operator, and their W-algebra duals," 
JHEP 1102 (2011) 114 [arXiv:1012.1355 [hep-th]]. 



\bibitem{SKAO}
J.~Shiraishi, H.~Kubo, H.~Awata and S.~Odake, ``A quantum deformation of the
  Virasoro algebra and the Macdonald symmetric functions,'' 
  Lett.\ Math.\ Phys.\ {\bf 38}, 33-51 (1996) [arXiv:q-alg/9507034].

\bibitem{FF95}
B.~Feigin and E.~Frenkel, ``Quantum $\mathcal{W}$-Algebras and Elliptic
  Algebras,'' Commun.\ Math.\ Phys.\ {\bf 178}, 653-678 (1996)
  [arXiv:q-alg/9508009].

\bibitem{AKOS95}
H.~Awata, H.~Kubo, S.~Odake and J.~Shiraishi, ``Quantum $\mathcal{W}_N$
  Algebras and Macdonald Polynomials,'' Commun.\ Math.\ Phys.\ {\bf 179},
  401-416 (1996) [arXiv:q-alg/9508011].

\bibitem{Waki1986}
M. Wakimoto, 
``Fock representations of the affine Lie algebra $A_1^{(1)}$," 
Comm. Math. Phys. 104, (1986), 605-609.


\bibitem{HJKO0005O}
Y. Hara, M. Jimbo, H. Konno, S. Odake and J. Shiraishi,
``On Lepowsky-Wilson's Z-algebra," 
Recent Developments in Infinite-Diminsional Lie Algebras and Conformal Field Theory, 
Proceedings of the Conference on Infinite-dimensional Lie Theory, Contemporary Mathematics 297, AMS, Providence,2002,  
[arXiv:math/0005203].

\bibitem{LW1981C}
J. Lepowsky and R. L. Wilson,
``Construction of the affine Lie algebra $A_1^{(1)}$,"
Commun. Math. Phys. {\bf 62} (1978) 43-53.

\bibitem{LW1981A}
J. Lepowsky and R. L. Wilson,
``A new family of algebras underlying the Rogers-Ramanujan 
 identities and generalizations," 
Proc. Natl. Acad. Sci. USA, {\bf 78} (1981) 7254-7258.


\bibitem{KZ}
V.G. Knizhnik and A.B. Zamolodchikov,
``Current algebra and Wess-Zumino model in two dimensions,"
Nucl. Phys. B {\bf 247}, (1984) 83-103.

\bibitem{EFK1997L}
P. I. Etingof, I. B. Frenkel and A. A. Kirillov, 
``{\it Lectures on Representation Theory and Knizhnik-Zamolodchikov Equations}," 
American Mathematical Society, 1997. 

\bibitem{Yama1011}
Y. Yamada, 
``A quantum isomonodromy equation and its application to 
 $\mathcal{N} = 2$ SU(N) gauge theories," 
 J. Phys. A {\bf 44} (2011) 055403 [arXiv:1011.0292 [hep-th]].

\end{thebibliography}
\small\baselineskip=.90\baselineskip
\let\bbb\bibitem\def\bibitem{\itemsep1pt\bbb}

%%%%%%%%%%%%%%%%%%%%%%END%%%%%%%%%%%%%%%%%%%%%%%
\end{document}